\documentclass[%reprint,
 amsmath,amssymb,
 aps,
 prl,
]{revtex4-2}

\usepackage{graphicx}% Include figure files
\usepackage{dcolumn}% Align table columns on decimal point
\usepackage{bm}% bold math
\usepackage[utf8]{inputenc}
\usepackage[letterpaper,margin=1in]{geometry}
\usepackage{commath}
\usepackage{amsmath}
\usepackage{amsfonts}
\usepackage{siunitx}
\usepackage{xcolor}

\newcommand{\ecoli}{\textit{Escherichia coli}}
\newcommand{\ec}{\textit{E.~coli}}
\newcommand{\supp}{Supplementary Material}

\newcommand{\mm}{Methods}
\newcommand{\sfg}{Fig.~S}

\newcommand{\smv}{Movie S}

\begin{document}
%\title{Geometry induced anomalies of active bacterial transport in structured environments}
   \title{Geometric effects induce anomalous size-dependent active transport in structured environments}
   %\title {Confinement enhances bacterial transport in structured environments}
   %\title{Geometry-induced anomalous size-dependence of bacterial diffusion in a micropillar array}
\author{Pooja Chopra}
\affiliation{Department of Physics, University of California, 5200 North Lake Road, Merced, CA 95343
}
\author{David Quint}
\affiliation{Department of Physics, University of California, 5200 North Lake Road, Merced, CA 95343
}
\author{Ajay Gopinathan}
\affiliation{Department of Physics, University of California, 5200 North Lake Road, Merced, CA 95343
}
\author{Bin Liu}
\affiliation{Department of Physics, University of California, 5200 North Lake Road, Merced, CA 95343
}

\begin{abstract}
Variations of transport efficiency in structured environments between distinct individuals in actively self-propelled systems is both hard to study and poorly understood. Here, we study the transport of a non-tumbling {\ecoli} strain, an active-matter archetype with intrinsic size variation but fairly uniform speed, through a periodic pillar array. We show that long-term transport switches from a trapping dominated state for shorter cells to a much more dispersive state for longer cells above a critical bacterial size set by the pillar array geometry. Using a combination of experiments and modeling, we show that this anomalous size-dependence arises from an enhancement of the escape rate from trapping for longer cells caused by nearby pillars. Our results show that geometric effects can lead to size being a sensitive tuning knob for transport in structured environments, with implications in general for active matter systems and, in particular, for the morphological adaptation of bacteria to structured habitats, spatial structuring of communities and for anti-biofouling materials design.
\end{abstract}
\maketitle

%\summary{Here, we study the transport of a non-tumbling {\ecoli} strain, an active-matter archetype with intrinsic size variation but fairly uniform speed, through a periodic pillar array. We show that the long-term transport of this bacterial system switches from a trapping dominated state for shorter cells to a much more dispersive state for longer cells above a critical particle size set by the pillar array geometry. %Specifically, we show that this enhanced dispersal for longer cells arises from its simultaneous interactions with multiple pillars, leading to an enhanced escape rate from trapping.Such geometric effects due to a lattice can lead to particle size being a sensitive tuning knob for transport in structured environments. Our results therefore have implications in general for active matter systems and, in particular, for bacterial adaptation to structured environments, spatial structuring of communities and for anti-biofouling materials design. 
%An active-matter system is known as a group of active particles or self-propelled agents that interact with surrounding environments and among themselves, including microorganisms and their artificial counterparts . }

%\section{Results}
%\subsection{Long-term transport of {\ec} in a micropillar array}
Structural features of environments have been recently shown to have a significant impact on the motility phases of active matter systems \cite{bechingerActiveParticlesComplex2016, morinDistortionDestructionColloidal2017, quintTopologicallyInducedSwarming2015,  chepizhkoDiffusionSubdiffusionTrapping2013, bertrandOptimizedDiffusionRunandTumble2018, sandorDynamicPhasesActive2017, pattanayakEnhancedDynamicsActive2019, phanBacterialRouteFinding2020, ribeiro_trapping_2020, yazdi_metamaterials_2020, brun-cosme-bruny_deflection_2020, reichhardt_directional_2020}.
However, much less is known about how the interplay between variations in individual particle geometry and environmental structure affects macroscopic transport. For motile bacterial systems, in particular, such effects \cite{wiolandFerromagneticAntiferromagneticOrder2016, creppyEffectMotilityTransport2019,  dehkharghaniBacterialScatteringMicrofluidic2019, makarchukEnhancedPropagationMotile2019, bhattacharjeeBacterialHoppingTrapping2019} can have implications for tunable transport in structured habitats. These implications arise because bacteria come in a variety of shapes and sizes across species \cite{youngSelectiveValueBacterial2006, rappeCultivationUbiquitousSAR112002, angertLargestBacterium1993} and even within a single strain \cite{hahnBacterialFilamentFormation1999, typasRegulationPeptidoglycanSynthesis2012, shenMorphologicalPlasticityBacteria2016}. 
It has been suggested that such widely distributed shapes and sizes are a consequence of adaptation to a diversity of features in their environments ranging from mechanical properties to nutrient availability \cite{youngSelectiveValueBacterial2006}. In particular, the optimization of transport or dispersal is known to provide a strong selective pressure for bacterial morphology evolution \cite{youngSelectiveValueBacterial2006, schuechMotileCurvedBacteria2019, liuHelicalMotionCell2014,persatCurvedShapeCaulobacter2014}. 
%For example, the curved shape of \textit{Caulobacter crescentus} has been independently attributed to the optimization of swimming efficiency in Newtonian fluids \cite{liuHelicalMotionCell2014} and also the enhancement of surface colonization in the presence of flow \cite{persatCurvedShapeCaulobacter2014}. 
For bacteria that live in structured or porous environments such as soil or tissue \cite{turnbullRoleBacterialMotility2001, balzanBacterialTranslocationOverview2007}, 
%it seems clear that the geometry of the surroundings should place strong transport constraints on bacterial size and shape.
%In the limit of strong confinement, the trapping of bacteria in voids can suppress overall transport \cite{bhattacharjeeBacterialHoppingTrapping2019} thereby potentially setting an upper bound on size. However, even away from this limit, it is to be noted that 
 proximity to a surface involves a whole host of physical interactions. These include hydrodynamic \cite{laugaSwimmingCirclesMotion2006, spagnolieGeometricCaptureEscape2015, shumModellingBacterialBehaviour2010, siposHydrodynamicTrappingSwimming2015}, electrostatic \cite{hermanssonDLVOTheoryMicrobial1999}, and steric \cite{drescherFluidDynamicsNoise2011} forces as well as flow induced effects \cite{dehkharghaniBacterialScatteringMicrofluidic2019, alonso-matillaTransportDispersionActive2019, creppyEffectMotilityTransport2019}, which could affect transport in a geometry dependent manner \cite{ tongDirectedMigrationMicroscale2018, makarchukEnhancedPropagationMotile2019, davieswykesGuidingMicroscaleSwimmers2017}.
%It has been shown, for instance, that hydrodynamic forces lead to the attraction or repulsion of a pusher swimmer (a force-dipole analogy of a typical swimming bacterium), depending on its geometry and orientation relative to the surfaces \cite{shumModellingBacterialBehaviour2010, spagnolieGeometricCaptureEscape2015, tongDirectedMigrationMicroscale2018, makarchukEnhancedPropagationMotile2019} and the curvature of the surface \cite{siposHydrodynamicTrappingSwimming2015, davieswykesGuidingMicroscaleSwimmers2017}. 
Here we study the possibility that the interplay between minor, intrinsic variations in the geometry of swimmers and structural features of the environment could lead to significant transport effects at the macroscopic scale.

One of the main challenges for such a study is that, in a standard microscope setting, a freely moving individual cell can only be observed, with adequate resolution of individual geometry, over a length scale similar to that of the cell size (${\sim}10$ $\mu$m). Meaningful statistics for its long-range transport over the millimeter scale is therefore hard to obtain. Here, we resolved this issue by following individual bacteria via a tracking microscope, where the microscope stage is adjusted in real-time to recenter the cell of interest in the field of view \cite{liuHelicalMotionCell2014}. 
To provide a structured environment, we fabricated a rectangular microfluidic channel, $1$ mm wide and 30 $\mu$m deep, embedded with square arrays of micropillars using a standard soft photolithography method (see {\mm}). A typical pillar array with pillar radii $R=15$ $\mu$m and lattice size $a=40$ $\mu$m (with the closest gap thus $d=a-2R=10$ $\mu$m) is shown in Fig.~1(a).  To focus on purely geometrical effects on transport that are applicable to generic active matter systems, we used a smooth swimming {\ecoli} strain, HCB437, which avoids any potential active response by the bacteria switching between run and tumble phases \cite{bergColiMotion2004}. Additionally, this strain shows a natural length variation from $2-10\mu$m between individuals allowing us to examine the effects of microscopic geometry on macroscopic transport.
We visualized the transport of these bacteria through the pillar array at a high ($60 \times$ or $100\times$) magnification over millimeters by reconstructing trajectories. This was done by stitching together single image frames during the course of tracking (see {\supp}) as shown in Fig.~1(b). Even though the tracked bacterium navigates the pillar array over a long distance, its detailed movement and orientation can still be resolved by visiting every single frame with a submicron resolution (Fig.~1(b)).  

We first examined in detail the trajectories of several bacteria with different sizes. Independent of size, bacteria are constrained to move within the open spaces between pillars and the presence of noise leads to an overall diffusive trajectory at long times. However, we noticed two qualitatively different modes of motility depending on size. Short cells ($2-5$ $\mu$m), on the one hand, frequent the pillar surfaces and move mostly in circular patterns, due to effective hydrodynamic trapping \cite{siposHydrodynamicTrappingSwimming2015, spagnolieGeometricCaptureEscape2015, tongDirectedMigrationMicroscale2018} by the pillar array (Fig.~2(a), also see {\smv}1). Distinct from a plane-wall-induced circulation \cite{laugaSwimmingCirclesMotion2006}, this circulation around the pillar is bidirectional, regardless of the chirality embedded in the flagellar filaments ({{\supp}}, Fig.~S7). Longer cells, on the other hand, appear to escape from such traps, resulting in more persistent movement along the directions of two orthogonal lattice vectors (here, $x$- and $y$-axes) (Fig.~2(a), also see {\smv}2). Thus, longer cells, despite feeling an increased confinement, showed an anomalous increase in their net transport.

To further quantify these distinct effects of the pillar array on bacteria with different sizes, we computed the probability distribution of  the centers of bacteria in space within a single unit cell. As shown in Fig.~2(b) and (c), shorter bacteria are concentrated near the pillar surface (consistent with a hydrodynamic attraction \cite{creppyEffectMotilityTransport2019}), while longer cells spend more time in the channels, confirming a size-dependent trapping effect. It is worth noting that tracking the front end of a cell provides essentially the same distribution near the pillar surface ({\supp}, {\sfg}5), suggesting a negligible role of cell length in volume exclusion.

%\subsection{Geometry-based model}

To understand the mechanism governing this effect, we considered the geometric constraints on a swimming bacterium due to neighboring pillars. For simplicity, the bacterium is regarded as a rod-shaped pusher of length $l_p$, which is the effective hydrodynamic size set by the flow profile of the entire swimmer including the cell body (of length $l$) and the flagella (Fig.~\ref{fig:geomodel}a, inset). Such a pusher tends to circulate around a single pillar in its natural state (without any neighboring pillars), due to the known hydrodynamic attraction between a solid surface and a generic pusher swimmer, including both bacteria and synthetic microswimmers \cite{berkeHydrodynamicAttractionSwimming2008, davieswykesGuidingMicroscaleSwimmers2017}. In the presence of the neighboring pillars, the allowable pusher sizes are restricted, with the maximum length $l_{p, \mathrm{max}}$ determined by the geometry (Fig.~\ref{fig:geomodel}a). In addition to the length constraint, a pusher is also subjected to hydrodynamic interactions from the neighboring pillars.  
Here, we consider a pusher circulating in the counter-clockwise direction and assume the pusher's orientation is always tangential to the pillar surface that it is circulating around.
Depending on the orientation of the pusher relative to the pillar lattice (denoted by angle $\gamma$), the pusher may experience a torque from the nearest-neighbor pillar that either promotes or inhibits its circulation around the pillar. Such distinct effects are determined by the orientation angle $\theta_p$ of the pusher relative to the surface normal of the nearest-neighbor pillar (Fig. 3a). Here, we consider only the normal component of the hydrodynamic force from the pillar, associated with the anisotropic drag coefficients in the presence of a nearby wall \cite{brennerSlowMotionSphere1961, laugaSwimmingCirclesMotion2006a}. For $\theta_p<0$, the normal force from the nearest-neighbor pillar provides a torque that tends to tip the pusher toward the center of the pillar it is circulating around, leading to an effective attraction to the pillar surface. For $\theta_p>0$, the torque due to the nearest-neighbor pillar tends to tip the pusher further away from the center causing the pusher to escape (likely along the tangent to the pillar surface). Considering the tangential components of the forces does not alter the directions of these torques, as long as the drag coefficient normal to the pillar surface dominates. It should be noted that we only treated the above pusher in a resistive-force-type manner to signify the geometric roles played by the nearest-neighbor pillar. More qualitative and quantitative insights of the hydrodynamic interactions requires resolving the flow field associated with a full pusher model including no-slip boundaries with pillar geometries \cite{spagnolieGeometricCaptureEscape2015}. 

This nearest-neighbor effect leads to a series of alternating attractive and repulsive zones along the perimeter of the pillar (with $\gamma\in [0, 2\pi)$ ), determined by the sign of $\theta_p$. Figure \ref{fig:geomodel}b shows the calculated $\theta_p$ and $l_{p, \mathrm{max}}$ (see {\supp}) for a counter-clockwise circulating pusher and a pillar lattice that is consistent with the experimental setting ($R/a=0.375$). The attractive ($\theta_p<0$) and repulsive ($\theta_p>0$) zones are shaded in red and blue, respectively. Zones that lack any constraints from the nearest-neighbor pillar (in white) are also considered attractive, due to the natural circulating state of pushers (in the absence of the neighboring pillars). This result shows four continuous attractive zones (blue plus white) along the perimeter of the pillar, with each spanning an arc angle $\Delta \gamma_0 = 1.1$ rad, centered near $\gamma = 0$, $\pi/2$, $\pi$, and $3\pi/2$, respectively. 
%The locations of these attractive zones are consistent with the peaks in the probability distribution of {\ec} near the pillar surfaces ({\supp}, {\sfg}2). 

To facilitate a more quantitative comparison between the experiment and the theoretical picture, we measured a residency arc angle $\Delta \gamma=\gamma_f - \gamma_i$, which is the difference between the two angles where the bacterium enters ($\gamma_i$) and escapes ($\gamma_f$) the vicinity of the pillar surface (Fig.~3c, inset). The bacteria continuously circulates the pillar over an arc subtending this angle without leaving the surface. The experimental residency arc angles $\Delta \gamma$, plotted against cell lengths $l$, are shown in Fig.~3c. At short cell lengths ($l\lesssim 4$ $\mu$m), $\Delta \gamma$ can span over larger angles ($\Delta \gamma > 2\pi$), corresponding to the presence of multi-turn circulations.
As $l$ increases, $\Delta \gamma$ becomes restricted only to small angles (e.g., $\Delta \gamma \lesssim \pi$ for $l\gtrsim 7$ $\mu$m). Such a restriction in the distribution of $\Delta \gamma$ is responsible for the decrease in its mean with increasing $l$. For sufficiently long $l$ ($l\gtrsim 10$ $\mu$m), the mean $\Delta \gamma$ falls beneath the size of the computed attractive zone ($\Delta \gamma_0=1.1$ rad), consistent with a highly constrained pusher that is unable to bypass any repulsive zones (as required for circulations beyond a single attractive zone). We also note that there is no such size dependence of $\Delta \gamma$ for sufficiently large gaps among pillars (\supp, Fig.~S6), which further validates our geometric arguments.   
%Delta \gamma$ span over a  the mean the distribution of $\Delta \gamma$ features a peak $\Delta \gamma_p \approx 1$ rad, which agrees surprisingly well with the size of the predicted attractive zones $\Delta \gamma_0= 1.1$ rad. To ensure that our geometric model captures the key effects of the pillar lattice on the pusher's circulation, we performed the same residency arc angle measurements for {\ec} in micropillar arrays with different lattice sizes $a$ and pillar radii $R$, and compared their peak arc angles $\Delta\gamma_p$ with the model predictions. Again, our results show a fairly good agreement between the experiments and theory (Fig.~3d). For given $l_{p, \mathrm{max}}$, we note that increasing confinement for larger $R/a$ leads to smaller residency arcs and suggesting less trapping and greater dispersal at long times.

%The above size-dependent resident arc angles $\Delta \gamma$ bypassing repulsive zones thus suggests the probability of escaping at repulsive zones $P_\mathrm{esc}$ as an indicator of the above geometric constraint.  

%To use the above geometric arguments for quantitative predictions, 
To further validate our model, we investigated the statistics of the angles, $\gamma_f$, at which bacteria escaped the pillar surface. The probability density function (PDF) of $\gamma_f$ shows peaks near the diagonal directions of the pillar lattice  (Fig.~4a), consistent with the predicted locations of the repulsive zones (Fig.~3b). This highly anisotropic distribution of $\gamma_f$ also justifies the use of a single escaping probability $P_\mathrm{esc}$ (within all repulsive zones only) for characterizing the size-dependent geometric effects. Such an escaping probability is consistent with the stochastic nature of the competition between both the hydrodynamic attraction from the orbited pillar and the ``repulsive'' contribution from the nearest-neighboring one, associated with the fluctuating orientation and location of a microswimmer. If we consider that bacteria only escape within repulsive zones, $P_\mathrm{esc}$ can be obtained as $P_\mathrm{esc}=\frac{N_e}{N_z}$, where $N_e$ and $N_z$ correspond respectively to the number of escaping events and the number of repulsive zones that bacteria cross during their travel along the pillars' perimeters. Noting that the angular separation between the centers of two adjacent repulsive zones is $\pi/2$ and neglecting the detailed escaping locations, the number $N_z$ is given by $N_z  = \sum_{i=1}^{N_e} \left([{2\Delta\gamma_i}/{\pi}]+1\right)$, where $i$ corresponds to the index of an escaping event and $[\cdot]$ denotes the integer part of a number. The escaping probability is thus the inverse of an ensemble average (denoted by $\left<\cdot\right>$), i.e.,  $P_\mathrm{esc}=\left<\left[{2\Delta \gamma}/{\pi}+1\right]\right>^{-1}$. We computed this probability $P_\mathrm{esc}$ from experiments for different groups of cell body lengths $l$. As shown in Fig.~4b, there is a transition from a low escaping probability ($P_\mathrm{exc}\approx 0.3$) for relatively shorter cells ($l\lesssim 5$ $\mu$m) to a high escaping probability ($P_\mathrm{exc}\approx 1$) for relatively longer ones ($l\gtrsim 7$ $\mu$m). A hyperbolic-tangent fit of the experimental data yields a function (solid curve in Fig.~4b) $P_\mathrm{esc}(l) = P_0+(1-P_0)\tanh\left({l-l_c}/{\Delta l}\right)$ with the critical cell body length $l_c = 6.0$ $\mu$m demarcating the distinct escaping behaviors.

To show how this geometric effect manifests itself in the long-time transport of bacteria, we simulated bacterial trajectories by a kinematic model (using the above $P_\mathrm{esc} (l)$): a bacterium that reaches a pillar surface stays on the surface and continues circulating the pillar if it is within an attractive zone or it escapes with a probability $P_\mathrm{esc}$ if it enters a repulsive zone (in red in Fig.~\ref{fig:geomodel}b). Here, we assumed that a bacterium moves at a constant speed $u$, which is a fairly good approximation for the non-tumbling mutant ({\supp}, {\sfg}8). After escaping along the direction tangential to the pillar's circumference, the bacterium moves at constant speed until it reaches an attractive or repulsive zone of the next pillar along its trajectory. A white noise in swimming direction was extracted from experimental trajectories ({\supp}, {\sfg}2) and introduced to the model swimmer when it is away from the pillar surface. We then computed the corresponding mean squared-displacement (MSD) for comparison with the experimental data (Fig.~4c). To secure optimal convergence, the MSD values were binned by path lengths $\Delta s$ and sampled over all trajectories within the same size $l$ group. For all experimental MSD (dots in Fig.~4c), cell body lengths $l$ ($3$ - $9$ $\mu$m) were grouped every $2$ $\mu$m to secure a sufficient number ($\gtrsim 20$) of long trajectories ($\gtrsim 200 $ $\mu$m) within each size category. All data points with less than 10 sampled trajectories were excluded. Each simulated MSD of the corresponding $l$ (dashed line in Fig.~4c) was computed over 100 model swimmers with each of them traveling $400 \times a$ in total path length $s$ from a random starting position. As shown in Fig.~4c, the simulated MSD reproduces well the characteristics of bacterial transport. In both experimental and simulated results, each MSD-$\Delta s$ curve contains a ballistic regime for small $\Delta s$ and a diffusive regime for large $\Delta s$. The size of the ballistic regime can be characterized by a ballistic length $\Delta s_c$ (Fig.~4c), set by the crossover point between two distinct scaling regimes for transport (with MSD exponents $\alpha=1$ and $\alpha=2$). Interestingly, the ballistic length $\Delta s_c/a \approx 1$ for relatively shorter cells, reminiscent of the effective reorientation time scale $\tau=a/u$ for point-like particles diffusing in obstacle networks \cite{jakuszeit_diffusion_2019, schakenraad_topotaxis_2020}.
While converging at short $\Delta s$, the long-time MSD-$\Delta s$ curves are noticeably higher for longer cells, corresponding to longer ballistic lengths $\Delta s_c$ at longer cell lengths $l$ and thus suggesting a finite-length effect. Again, longer cells, despite feeling an increased confinement, display an anomalous increase in their MSD due to the geometric effects of the neighboring pillars. These quantitative agreements between the experiment and theory further confirms the dominant role played by geometry. 

%\section{Discussion}
Our study illustrates the role of individual morphology in the transport of active particles in general and bacteria, in particular, through periodic structured media. 
%We showed that a relatively longer agent disperses with a {\it higher} diffusivity, due to a lattice effect. This effect arises from the increased likelihood of a simultaneous interaction with multiple pillars for longer agents whereas hydrodynamic attraction to a single pillar dominates for shorter counterparts. Our work thus demonstrates the potential for exploiting lattice geometry to sort active particles and tune their collective motion based on individual morphology with high sensitivity \cite{longCellcellCommunicationEnhances2017,  nishiguchiEngineeringBacterialVortex2018}. 
In living systems, this surprising enhancement of transport for longer cells, combined with a maximal cell size set by interstitial spaces of the lattice, suggests an optimal cell size potentially determined by the geometry of a porous environment \cite{bakkenRelationshipCellSize1987, youngSelectiveValueBacterial2006,  schuechMotileCurvedBacteria2019}. Similar geometry-sensitive effects in transport of wild-type strains may already be present, but not explicitly identified, in other studies of bacterial transport \cite{bhattacharjeeBacterialHoppingTrapping2019}. Generalizing our geometric effects to such 3D environments will enable targeted design of environmental geometry for desired size-dependent transport and collective motion \cite{ wiolandFerromagneticAntiferromagneticOrder2016, longCellcellCommunicationEnhances2017, nishiguchiEngineeringBacterialVortex2018}. For a non-periodic lattice, it is expected that the above geometric effect still applies in general, since any geometric constraints due to nearest-neighbor pillars always tend to influence the longer cells first before they can influence shorter cells. However, the location and size of the repulsive zones, as well as the critical cell-body length now vary for each pillar, leading to more complex escaping zones and thus more complex global cell kinematics. Also, in the extreme case that multiple pillars are within the vicinity, a long cell will be more easily jammed as it requires more room for reorientation, contributing to another type of geometric constraint. These potential effects due to disorder in the crystalline structure of the environment will be investigated in our future work.

Our findings also indicate that the transition between localized and dispersive modes is sharp and occurs at a critical value of bacterial size, controlled by the porous environment. Bacteria with typical sizes near this critical value may be able to access both modes of transport by adaptive change in their size based on the local nutrient conditions or other desired transport needs.% For example, {\ec} undergoes filamentous growth under starvation condition \cite{wainwrightMorphologicalChangesIncluding1999}. The higher diffusivities for longer cells that we show, may thus allow for escape from a structured environment by filamentation triggered by the depletion of local nutrients while trapped. 
This suggests, for example, an unexplored benefit of filamentation under starvation conditions in {\ec}, in addition to others that have been proposed in the literature \cite{wainwrightMorphologicalChangesIncluding1999, millerSOSResponseInduction2004, justiceFilamentationEscherichiaColi2006}.
%We also expect the geometric principle here to be applicable to other types of active-matter systems, including active Brownian particles that are capable of active reorientation, such as the wild-type {\ec} strain.
 %Such active reorientation can increase the escaping probability of shorter trapped cells while it is also possible for neighboring pillars to potentially restrict the active reorientation of longer cells and thus suppress escape. Similar geometry-sensitive effects in transport of wild-type strains may already be present, but not explicitly identified, in other studies of bacterial transport \cite{bhattacharjeeBacterialHoppingTrapping2019}. Generalizing our geometric effects to such 3D environments will enable targeted design of environmental geometry for desired size-dependent transport and collective motion \cite{ wiolandFerromagneticAntiferromagneticOrder2016, longCellcellCommunicationEnhances2017, nishiguchiEngineeringBacterialVortex2018}. 
 Our results also have implications for the spatial structure of naturally occurring bacterial colonies in structured environments where spatial location within the colony could be correlated with age dependent cell-size, due to differential transport.

%The fact that the pattern of repulsive and attractive zones around the pillar are tunable by the lattice geometry enables richer applications beyond our current studies. 
In nature, patterned structures with periodic lattices are widely found on antibiofouling surfaces, such as cicada wings \cite{ivanovaNaturalBactericidalSurfaces2012} and shark skins \cite{schumacherEngineeredAntifoulingMicrotopographies2007}. The geometric effects we have identified thus provide a new perspective for revisiting these microscale structures in relation to their antibiofouling effects.
%So far, these antibiofouling effects have been attributed to the mechanical rupture of cell walls with a single nano- and micro-scale spike \cite{dicksonNanopatternedPolymerSurfaces2015}. Whether these structures perform any collective role is still not fully understood. The geometric effects we have identified thus provide a new perspective for revisiting these microscale structures in relation to their antibiofouling effects.
Conversely, our work also suggests ways to engineer surfaces so as to either increase or decrease the residency of different bacterial strains with slightly different sizes or even differentiating age structured populations, which may also be of interest in biofouling applications. Such ideas could also be applied to designing environments for the desired sorting or guiding of synthetic microswimmers as well as the geometric design of individual swimmers. 

%\section{Acknowledgements}
\begin{acknowledgements}
We would like to thank Howard Berg for providing the {\ec} mutant strain used in this study. We also acknowledge Jeremias Gonzalez, Yu Zeng, and Jacinta Conrad for insightful discussions. This work was supported by National Science Foundation Grant CBET-2046822, CBET-1706511, and NSF-CREST: Center for Cellular and Bio-molecular Machines (CCBM) at UC Merced (HRD-1547848). B.L. thanks the support of Hellman Foundation. A.G. also acknowledges support from National Science Foundation (NSF) grant DMS-1616926, partial support from the NSF Center for Engineering Mechanobiology grant NSF CMMI-154857 and the hospitality of the Aspen Center for Physics, which is supported by NSF grant PHY-1607611.
\end{acknowledgements}

%\section*{Contribution}
%B.L and A.G. conceptualized the study. P.C., D.Q., A.G., and B.L. designed the experiments. P.C.,B.L. and A.G. performed the simulations. P.C. carried out imaging experiments, and P.C. and D.Q. analyzed the data. P.C., A.G. and B.L. wrote the paper. All authors reviewed and edited the paper prior to submission. 

\bibliographystyle{naturemag}

\begin{thebibliography}{10}
\expandafter\ifx\csname url\endcsname\relax
  \def\url#1{\texttt{#1}}\fi
\expandafter\ifx\csname urlprefix\endcsname\relax\def\urlprefix{URL }\fi
\providecommand{\bibinfo}[2]{#2}
\providecommand{\eprint}[2][]{\url{#2}}

\bibitem{bechingerActiveParticlesComplex2016}
\bibinfo{author}{Bechinger, C.} \emph{et~al.}
\newblock \bibinfo{title}{Active particles in complex and crowded
  environments}.
\newblock \emph{\bibinfo{journal}{Reviews of Modern Physics}}
  \textbf{\bibinfo{volume}{88}}, \bibinfo{pages}{045006}
  (\bibinfo{year}{2016}).

\bibitem{morinDistortionDestructionColloidal2017}
\bibinfo{author}{Morin, A.}, \bibinfo{author}{Desreumaux, N.},
  \bibinfo{author}{Caussin, J.-B.} \& \bibinfo{author}{Bartolo, D.}
\newblock \bibinfo{title}{Distortion and destruction of colloidal flocks in
  disordered environments}.
\newblock \emph{\bibinfo{journal}{Nature Physics}}
  \textbf{\bibinfo{volume}{13}}, \bibinfo{pages}{63--67}
  (\bibinfo{year}{2017}).

\bibitem{quintTopologicallyInducedSwarming2015}
\bibinfo{author}{Quint, D.~A.} \& \bibinfo{author}{Gopinathan, A.}
\newblock \bibinfo{title}{Topologically induced swarming phase transition on a
  {{2D}} percolated lattice}.
\newblock \emph{\bibinfo{journal}{Physical Biology}}
  \textbf{\bibinfo{volume}{12}}, \bibinfo{pages}{046008}
  (\bibinfo{year}{2015}).

\bibitem{chepizhkoDiffusionSubdiffusionTrapping2013}
\bibinfo{author}{Chepizhko, O.} \& \bibinfo{author}{Peruani, F.}
\newblock \bibinfo{title}{Diffusion, {{Subdiffusion}}, and {{Trapping}} of
  {{Active Particles}} in {{Heterogeneous Media}}}.
\newblock \emph{\bibinfo{journal}{Physical Review Letters}}
  \textbf{\bibinfo{volume}{111}}, \bibinfo{pages}{160604}
  (\bibinfo{year}{2013}).

\bibitem{bertrandOptimizedDiffusionRunandTumble2018}
\bibinfo{author}{Bertrand, T.}, \bibinfo{author}{Zhao, Y.},
  \bibinfo{author}{B{\'e}nichou, O.}, \bibinfo{author}{Tailleur, J.} \&
  \bibinfo{author}{Voituriez, R.}
\newblock \bibinfo{title}{Optimized {{Diffusion}} of {{Run}}-and-{{Tumble
  Particles}} in {{Crowded Environments}}}.
\newblock \emph{\bibinfo{journal}{Physical Review Letters}}
  \textbf{\bibinfo{volume}{120}}, \bibinfo{pages}{198103}
  (\bibinfo{year}{2018}).

\bibitem{sandorDynamicPhasesActive2017}
\bibinfo{author}{S{\'a}ndor, C.}, \bibinfo{author}{Lib{\'a}l, A.},
  \bibinfo{author}{Reichhardt, C.} \& \bibinfo{author}{Olson~Reichhardt, C.~J.}
\newblock \bibinfo{title}{Dynamic phases of active matter systems with quenched
  disorder}.
\newblock \emph{\bibinfo{journal}{Physical Review E}}
  \textbf{\bibinfo{volume}{95}}, \bibinfo{pages}{032606}
  (\bibinfo{year}{2017}).

\bibitem{pattanayakEnhancedDynamicsActive2019}
\bibinfo{author}{Pattanayak, S.}, \bibinfo{author}{Das, R.},
  \bibinfo{author}{Kumar, M.} \& \bibinfo{author}{Mishra, S.}
\newblock \bibinfo{title}{Enhanced dynamics of active {{Brownian}} particles in
  periodic obstacle arrays and corrugated channels}.
\newblock \emph{\bibinfo{journal}{The European Physical Journal E}}
  \textbf{\bibinfo{volume}{42}}, \bibinfo{pages}{62} (\bibinfo{year}{2019}).

\bibitem{phanBacterialRouteFinding2020}
\bibinfo{author}{Phan, T.~V.} \emph{et~al.}
\newblock \bibinfo{title}{Bacterial route finding and collective escape in
  mazes and fractals}.
\newblock \emph{\bibinfo{journal}{Physical Review X}}
  \textbf{\bibinfo{volume}{10}}, \bibinfo{pages}{031017}
  (\bibinfo{year}{2020}).

\bibitem{ribeiro_trapping_2020}
\bibinfo{author}{Ribeiro, H.~E.}, \bibinfo{author}{Ferreira, W.~P.} \&
  \bibinfo{author}{Potiguar, F.~Q.}
\newblock \bibinfo{title}{Trapping and sorting of active matter in a periodic
  background potential}.
\newblock \emph{\bibinfo{journal}{Physical Review E}}
  \textbf{\bibinfo{volume}{101}}, \bibinfo{pages}{032126}
  (\bibinfo{year}{2020}).

\bibitem{yazdi_metamaterials_2020}
\bibinfo{author}{Yazdi, S.}, \bibinfo{author}{Aragones, J.~L.},
  \bibinfo{author}{Coulter, J.} \& \bibinfo{author}{Alexander-Katz, A.}
\newblock \bibinfo{title}{Metamaterials for {Active} {Colloid} {Transport}}
  (\bibinfo{year}{2020}).
\newblock \bibinfo{note}{Publisher: arXiv Version Number: 1}.

\bibitem{brun-cosme-bruny_deflection_2020}
\bibinfo{author}{Brun-Cosme-Bruny, M.} \emph{et~al.}
\newblock \bibinfo{title}{Deflection of phototactic microswimmers through
  obstacle arrays}.
\newblock \emph{\bibinfo{journal}{Physical Review Fluids}}
  \textbf{\bibinfo{volume}{5}}, \bibinfo{pages}{093302} (\bibinfo{year}{2020}).

\bibitem{reichhardt_directional_2020}
\bibinfo{author}{Reichhardt, C.} \& \bibinfo{author}{Reichhardt, C. J.~O.}
\newblock \bibinfo{title}{Directional locking effects for active matter
  particles coupled to a periodic substrate}.
\newblock \emph{\bibinfo{journal}{Physical Review E}}
  \textbf{\bibinfo{volume}{102}}, \bibinfo{pages}{042616}
  (\bibinfo{year}{2020}).

\bibitem{wiolandFerromagneticAntiferromagneticOrder2016}
\bibinfo{author}{Wioland, H.}, \bibinfo{author}{Woodhouse, F.~G.},
  \bibinfo{author}{Dunkel, J.} \& \bibinfo{author}{Goldstein, R.~E.}
\newblock \bibinfo{title}{Ferromagnetic and antiferromagnetic order in bcterial
  vortex lattices}.
\newblock \emph{\bibinfo{journal}{Nature Physics}}
  \textbf{\bibinfo{volume}{12}}, \bibinfo{pages}{341--345}
  (\bibinfo{year}{2016}).

\bibitem{creppyEffectMotilityTransport2019}
\bibinfo{author}{Creppy, A.}, \bibinfo{author}{Cl{\'e}ment, E.},
  \bibinfo{author}{Douarche, C.}, \bibinfo{author}{D'Angelo, M.~V.} \&
  \bibinfo{author}{Auradou, H.}
\newblock \bibinfo{title}{Effect of motility on the transport of bacteria
  populations through a porous medium}.
\newblock \emph{\bibinfo{journal}{Physical Review Fluids}}
  \textbf{\bibinfo{volume}{4}}, \bibinfo{pages}{013102} (\bibinfo{year}{2019}).

\bibitem{dehkharghaniBacterialScatteringMicrofluidic2019}
\bibinfo{author}{Dehkharghani, A.}, \bibinfo{author}{Waisbord, N.},
  \bibinfo{author}{Dunkel, J.} \& \bibinfo{author}{Guasto, J.~S.}
\newblock \bibinfo{title}{Bacterial scattering in microfluidic crystal flows
  reveals giant active {{Taylor}}\textendash{{Aris}} dispersion}.
\newblock \emph{\bibinfo{journal}{Proceedings of the National Academy of
  Sciences}} \textbf{\bibinfo{volume}{116}}, \bibinfo{pages}{11119--11124}
  (\bibinfo{year}{2019}).

\bibitem{makarchukEnhancedPropagationMotile2019}
\bibinfo{author}{Makarchuk, S.}, \bibinfo{author}{Braz, V.~C.},
  \bibinfo{author}{Ara{\'u}jo, N. A.~M.}, \bibinfo{author}{Ciric, L.} \&
  \bibinfo{author}{Volpe, G.}
\newblock \bibinfo{title}{Enhanced propagation of motile bacteria on surfaces
  due to forward scattering}.
\newblock \emph{\bibinfo{journal}{Nature Communications}}
  \textbf{\bibinfo{volume}{10}}, \bibinfo{pages}{4110} (\bibinfo{year}{2019}).

\bibitem{bhattacharjeeBacterialHoppingTrapping2019}
\bibinfo{author}{Bhattacharjee, T.} \& \bibinfo{author}{Datta, S.~S.}
\newblock \bibinfo{title}{Bacterial hopping and trapping in porous media}.
\newblock \emph{\bibinfo{journal}{Nature Communications}}
  \textbf{\bibinfo{volume}{10}}, \bibinfo{pages}{2075} (\bibinfo{year}{2019}).

\bibitem{youngSelectiveValueBacterial2006}
\bibinfo{author}{Young, K.~D.}
\newblock \bibinfo{title}{The selective value of bacterial shape}.
\newblock \emph{\bibinfo{journal}{Microbiology and Molecular Biology Reviews}}
  \textbf{\bibinfo{volume}{70}}, \bibinfo{pages}{660--703}
  (\bibinfo{year}{2006}).

\bibitem{rappeCultivationUbiquitousSAR112002}
\bibinfo{author}{Rapp{\'e}, M.~S.}, \bibinfo{author}{Connon, S.~A.},
  \bibinfo{author}{Vergin, K.~L.} \& \bibinfo{author}{Giovannoni, S.~J.}
\newblock \bibinfo{title}{Cultivation of the ubiquitous {{SAR11}} marine
  bacterioplankton clade}.
\newblock \emph{\bibinfo{journal}{Nature}} \textbf{\bibinfo{volume}{418}},
  \bibinfo{pages}{630--633} (\bibinfo{year}{2002}).

\bibitem{angertLargestBacterium1993}
\bibinfo{author}{Angert, E.~R.}, \bibinfo{author}{Clements, K.~D.} \&
  \bibinfo{author}{Pace, N.~R.}
\newblock \bibinfo{title}{The largest bacterium}.
\newblock \emph{\bibinfo{journal}{Nature}} \textbf{\bibinfo{volume}{362}},
  \bibinfo{pages}{239--241} (\bibinfo{year}{1993}).

\bibitem{hahnBacterialFilamentFormation1999}
\bibinfo{author}{Hahn, M.~W.}, \bibinfo{author}{Moore, E.~R.} \&
  \bibinfo{author}{H{\"o}fle, M.~G.}
\newblock \bibinfo{title}{Bacterial filament formation, a defense mechanism
  against flagellate grazing, is growth rate controlled in bacteria of
  different phyla}.
\newblock \emph{\bibinfo{journal}{Applied and Environmental Microbiology}}
  \textbf{\bibinfo{volume}{65}}, \bibinfo{pages}{25--35}
  (\bibinfo{year}{1999}).

\bibitem{typasRegulationPeptidoglycanSynthesis2012}
\bibinfo{author}{Typas, A.}, \bibinfo{author}{Banzhaf, M.},
  \bibinfo{author}{Gross, C.~A.} \& \bibinfo{author}{Vollmer, W.}
\newblock \bibinfo{title}{From the regulation of peptidoglycan synthesis to
  bacterial growth and morphology}.
\newblock \emph{\bibinfo{journal}{Nature Reviews Microbiology}}
  \textbf{\bibinfo{volume}{10}}, \bibinfo{pages}{123--136}
  (\bibinfo{year}{2012}).

\bibitem{shenMorphologicalPlasticityBacteria2016}
\bibinfo{author}{Shen, J.-P.} \& \bibinfo{author}{Chou, C.-F.}
\newblock \bibinfo{title}{Morphological plasticity of
  bacteria\textemdash{{Open}} questions}.
\newblock \emph{\bibinfo{journal}{Biomicrofluidics}}
  \textbf{\bibinfo{volume}{10}}, \bibinfo{pages}{031501}
  (\bibinfo{year}{2016}).

\bibitem{schuechMotileCurvedBacteria2019}
\bibinfo{author}{Schuech, R.}, \bibinfo{author}{Hoehfurtner, T.},
  \bibinfo{author}{Smith, D.~J.} \& \bibinfo{author}{Humphries, S.}
\newblock \bibinfo{title}{Motile curved bacteria are {{Pareto}}-optimal}.
\newblock \emph{\bibinfo{journal}{Proceedings of the National Academy of
  Sciences}} \textbf{\bibinfo{volume}{116}}, \bibinfo{pages}{14440--14447}
  (\bibinfo{year}{2019}).

\bibitem{liuHelicalMotionCell2014}
\bibinfo{author}{Liu, B.} \emph{et~al.}
\newblock \bibinfo{title}{Helical motion of the cell body enhances
  {{{\emph{Caulobacter}}}}{\emph{ crescentus}} motility}.
\newblock \emph{\bibinfo{journal}{Proceedings of the National Academy of
  Sciences}} \textbf{\bibinfo{volume}{111}}, \bibinfo{pages}{11252--11256}
  (\bibinfo{year}{2014}).

\bibitem{persatCurvedShapeCaulobacter2014}
\bibinfo{author}{Persat, A.}, \bibinfo{author}{Stone, H.~A.} \&
  \bibinfo{author}{Gitai, Z.}
\newblock \bibinfo{title}{The curved shape of {{{\emph{Caulobacter}}}}{\emph{
  crescentus}} enhances surface colonization in flow}.
\newblock \emph{\bibinfo{journal}{Nature Communications}}
  \textbf{\bibinfo{volume}{5}}, \bibinfo{pages}{12855} (\bibinfo{year}{2014}).

\bibitem{turnbullRoleBacterialMotility2001}
\bibinfo{author}{Turnbull, G.~A.}, \bibinfo{author}{Morgan, J.~W.},
  \bibinfo{author}{Whipps, J.~M.} \& \bibinfo{author}{Saunders, J.~R.}
\newblock \bibinfo{title}{The role of bacterial motility in the survival and
  spread of {{Pseudomonas}} fluorescens in soil and in the attachment and
  colonisation of wheat roots}.
\newblock \emph{\bibinfo{journal}{FEMS Microbiology Ecology}}
  \textbf{\bibinfo{volume}{36}}, \bibinfo{pages}{21--31}
  (\bibinfo{year}{2001}).

\bibitem{balzanBacterialTranslocationOverview2007}
\bibinfo{author}{Balzan, S.}, \bibinfo{author}{{de Almeida Quadros}, C.},
  \bibinfo{author}{{de Cleva}, R.}, \bibinfo{author}{Zilberstein, B.} \&
  \bibinfo{author}{Cecconello, I.}
\newblock \bibinfo{title}{Bacterial translocation: {{Overview}} of mechanisms
  and clinical impact}.
\newblock \emph{\bibinfo{journal}{Journal of Gastroenterology and Hepatology}}
  \textbf{\bibinfo{volume}{22}}, \bibinfo{pages}{464--471}
  (\bibinfo{year}{2007}).

\bibitem{laugaSwimmingCirclesMotion2006}
\bibinfo{author}{Lauga, E.}, \bibinfo{author}{DiLuzio, W.~R.},
  \bibinfo{author}{Whitesides, G.~M.} \& \bibinfo{author}{Stone, H.~A.}
\newblock \bibinfo{title}{Swimming in circles: Motion of bacteria near solid
  boundaries}.
\newblock \emph{\bibinfo{journal}{Biophysical Journal}}
  \textbf{\bibinfo{volume}{90}}, \bibinfo{pages}{400--412}
  (\bibinfo{year}{2006}).

\bibitem{spagnolieGeometricCaptureEscape2015}
\bibinfo{author}{Spagnolie, S.~E.}, \bibinfo{author}{{Moreno-Flores}, G.~R.},
  \bibinfo{author}{Bartolo, D.} \& \bibinfo{author}{Lauga, E.}
\newblock \bibinfo{title}{Geometric capture and escape of a microswimmer
  colliding with an obstacle}.
\newblock \emph{\bibinfo{journal}{Soft Matter}} \textbf{\bibinfo{volume}{11}},
  \bibinfo{pages}{3396--3411} (\bibinfo{year}{2015}).

\bibitem{shumModellingBacterialBehaviour2010}
\bibinfo{author}{Shum, H.}, \bibinfo{author}{Gaffney, E.~A.} \&
  \bibinfo{author}{Smith, D.~J.}
\newblock \bibinfo{title}{Modelling bacterial behaviour close to a no-slip
  plane boundary: The influence of bacterial geometry}.
\newblock \emph{\bibinfo{journal}{Proceedings of the Royal Society A:
  Mathematical, Physical and Engineering Sciences}}
  \textbf{\bibinfo{volume}{466}}, \bibinfo{pages}{1725--1748}
  (\bibinfo{year}{2010}).

\bibitem{siposHydrodynamicTrappingSwimming2015}
\bibinfo{author}{Sipos, O.}, \bibinfo{author}{Nagy, K.},
  \bibinfo{author}{Di~Leonardo, R.} \& \bibinfo{author}{Galajda, P.}
\newblock \bibinfo{title}{Hydrodynamic trapping of swimming bacteria by convex
  walls}.
\newblock \emph{\bibinfo{journal}{Physical Review Letters}}
  \textbf{\bibinfo{volume}{114}}, \bibinfo{pages}{258104}
  (\bibinfo{year}{2015}).

\bibitem{hermanssonDLVOTheoryMicrobial1999}
\bibinfo{author}{Hermansson, M.}
\newblock \bibinfo{title}{The {{DLVO}} theory in microbial adhesion}.
\newblock \emph{\bibinfo{journal}{Colloids and Surfaces B: Biointerfaces}}
  \textbf{\bibinfo{volume}{14}}, \bibinfo{pages}{105--119}
  (\bibinfo{year}{1999}).

\bibitem{drescherFluidDynamicsNoise2011}
\bibinfo{author}{Drescher, K.}, \bibinfo{author}{Dunkel, J.},
  \bibinfo{author}{Cisneros, L.~H.}, \bibinfo{author}{Ganguly, S.} \&
  \bibinfo{author}{Goldstein, R.~E.}
\newblock \bibinfo{title}{Fluid dynamics and noise in bacterial cell-cell and
  cell-surface scattering}.
\newblock \emph{\bibinfo{journal}{Proceedings of the National Academy of
  Sciences}} \textbf{\bibinfo{volume}{108}}, \bibinfo{pages}{10940--10945}
  (\bibinfo{year}{2011}).

\bibitem{alonso-matillaTransportDispersionActive2019}
\bibinfo{author}{{Alonso-Matilla}, R.}, \bibinfo{author}{Chakrabarti, B.} \&
  \bibinfo{author}{Saintillan, D.}
\newblock \bibinfo{title}{Transport and dispersion of active particles in
  periodic porous media}.
\newblock \emph{\bibinfo{journal}{Physical Review Fluids}}
  \textbf{\bibinfo{volume}{4}}, \bibinfo{pages}{043101} (\bibinfo{year}{2019}).

\bibitem{tongDirectedMigrationMicroscale2018}
\bibinfo{author}{Tong, J.} \& \bibinfo{author}{Shelley, M.~J.}
\newblock \bibinfo{title}{Directed migration of microscale swimmers by an array
  of shaped obstacles: modeling and shape optimization}.
\newblock \emph{\bibinfo{journal}{SIAM Journal on Applied Mathematics}}
  \textbf{\bibinfo{volume}{78}}, \bibinfo{pages}{2370--2392}
  (\bibinfo{year}{2018}).

\bibitem{davieswykesGuidingMicroscaleSwimmers2017}
\bibinfo{author}{Davies~Wykes, M.~S.} \emph{et~al.}
\newblock \bibinfo{title}{Guiding microscale swimmers using teardrop-shaped
  posts}.
\newblock \emph{\bibinfo{journal}{Soft Matter}} \textbf{\bibinfo{volume}{13}},
  \bibinfo{pages}{4681--4688} (\bibinfo{year}{2017}).

\bibitem{bergColiMotion2004}
\bibinfo{editor}{Berg, H.~C.} (ed.) \emph{\bibinfo{title}{E. Coli in
  {{Motion}}}}.
\newblock Biological and {{Medical Physics}}, {{Biomedical Engineering}}
  (\bibinfo{publisher}{{Springer New York}}, \bibinfo{address}{{New York, NY}},
  \bibinfo{year}{2004}).

\bibitem{berkeHydrodynamicAttractionSwimming2008}
\bibinfo{author}{Berke, A.~P.}, \bibinfo{author}{Turner, L.},
  \bibinfo{author}{Berg, H.~C.} \& \bibinfo{author}{Lauga, E.}
\newblock \bibinfo{title}{Hydrodynamic {{Attraction}} of {{Swimming
  Microorganisms}} by {{Surfaces}}}.
\newblock \emph{\bibinfo{journal}{Physical Review Letters}}
  \textbf{\bibinfo{volume}{101}}, \bibinfo{pages}{038102}
  (\bibinfo{year}{2008}).

\bibitem{brennerSlowMotionSphere1961}
\bibinfo{author}{Brenner, H.}
\newblock \bibinfo{title}{The slow motion of a sphere through a viscous fluid
  towards a plane surface}.
\newblock \emph{\bibinfo{journal}{Chemical Engineering Science}}
  \textbf{\bibinfo{volume}{16}}, \bibinfo{pages}{242--251}
  (\bibinfo{year}{1961}).

\bibitem{laugaSwimmingCirclesMotion2006a}
\bibinfo{author}{Lauga, E.}, \bibinfo{author}{DiLuzio, W.~R.},
  \bibinfo{author}{Whitesides, G.~M.} \& \bibinfo{author}{Stone, H.~A.}
\newblock \bibinfo{title}{Swimming in circles: Motion of bacteria near solid
  boundaries}.
\newblock \emph{\bibinfo{journal}{Biophysical Journal}}
  \textbf{\bibinfo{volume}{90}}, \bibinfo{pages}{400--412}
  (\bibinfo{year}{2006}).

\bibitem{jakuszeit_diffusion_2019}
\bibinfo{author}{Jakuszeit, T.}, \bibinfo{author}{Croze, O.~A.} \&
  \bibinfo{author}{Bell, S.}
\newblock \bibinfo{title}{Diffusion of active particles in a complex
  environment: {Role} of surface scattering}.
\newblock \emph{\bibinfo{journal}{Physical Review E}}
  \textbf{\bibinfo{volume}{99}}, \bibinfo{pages}{012610}
  (\bibinfo{year}{2019}).

\bibitem{schakenraad_topotaxis_2020}
\bibinfo{author}{Schakenraad, K.} \emph{et~al.}
\newblock \bibinfo{title}{Topotaxis of active {Brownian} particles}.
\newblock \emph{\bibinfo{journal}{Physical Review E}}
  \textbf{\bibinfo{volume}{101}}, \bibinfo{pages}{032602}
  (\bibinfo{year}{2020}).

\bibitem{bakkenRelationshipCellSize1987}
\bibinfo{author}{Bakken, L.~R.} \& \bibinfo{author}{Olsen, R.~A.}
\newblock \bibinfo{title}{The relationship between cell size and viability of
  soil bacteria}.
\newblock \emph{\bibinfo{journal}{Microbial Ecology}}
  \textbf{\bibinfo{volume}{13}}, \bibinfo{pages}{103--114}
  (\bibinfo{year}{1987}).

\bibitem{longCellcellCommunicationEnhances2017}
\bibinfo{author}{Long, Z.}, \bibinfo{author}{Quaife, B.},
  \bibinfo{author}{Salman, H.} \& \bibinfo{author}{Oltvai, Z.~N.}
\newblock \bibinfo{title}{Cell-cell communication enhances bacterial chemotaxis
  toward external attractants}.
\newblock \emph{\bibinfo{journal}{Scientific Reports}}
  \textbf{\bibinfo{volume}{7}}, \bibinfo{pages}{12855} (\bibinfo{year}{2017}).

\bibitem{nishiguchiEngineeringBacterialVortex2018}
\bibinfo{author}{Nishiguchi, D.}, \bibinfo{author}{Aranson, I.~S.},
  \bibinfo{author}{Snezhko, A.} \& \bibinfo{author}{Sokolov, A.}
\newblock \bibinfo{title}{Engineering bacterial vortex lattice via direct laser
  lithography}.
\newblock \emph{\bibinfo{journal}{Nature Communications}}
  \textbf{\bibinfo{volume}{9}}, \bibinfo{pages}{4486} (\bibinfo{year}{2018}).

\bibitem{wainwrightMorphologicalChangesIncluding1999}
\bibinfo{author}{Wainwright, M.}, \bibinfo{author}{Canham, L.~T.},
  \bibinfo{author}{{Al-Wajeeh}, K.} \& \bibinfo{author}{Reeves, C.~L.}
\newblock \bibinfo{title}{Morphological changes (including filamentation) in
  {$E$\emph{scherichia coli}} grown under starvation conditions on silicon
  wafers and other surfaces}.
\newblock \emph{\bibinfo{journal}{Letters in Applied Microbiology}}
  \textbf{\bibinfo{volume}{29}}, \bibinfo{pages}{224--227}
  (\bibinfo{year}{1999}).

\bibitem{millerSOSResponseInduction2004}
\bibinfo{author}{Miller, C.} \emph{et~al.}
\newblock \bibinfo{title}{{{SOS}} response induction by {$\beta$}-lactams and
  bacterial defense against antibiotic lethality}.
\newblock \emph{\bibinfo{journal}{Science}} \textbf{\bibinfo{volume}{305}},
  \bibinfo{pages}{1629--1631} (\bibinfo{year}{2004}).

\bibitem{justiceFilamentationEscherichiaColi2006}
\bibinfo{author}{Justice, S.~S.}, \bibinfo{author}{Hunstad, D.~A.},
  \bibinfo{author}{Seed, P.~C.} \& \bibinfo{author}{Hultgren, S.~J.}
\newblock \bibinfo{title}{Filamentation by {$E$\emph{scherichia coli}} subverts
  innate defenses during urinary tract infection}.
\newblock \emph{\bibinfo{journal}{Proceedings of the National Academy of
  Sciences}} \textbf{\bibinfo{volume}{103}}, \bibinfo{pages}{19884--19889}
  (\bibinfo{year}{2006}).

\bibitem{ivanovaNaturalBactericidalSurfaces2012}
\bibinfo{author}{Ivanova, E.~P.} \emph{et~al.}
\newblock \bibinfo{title}{Natural bactericidal surfaces: mechanical rupture of
  {{{\emph{Pseudomonas}}}}{\emph{ aeruginosa}} cells by cicada wings}.
\newblock \emph{\bibinfo{journal}{Small}} \textbf{\bibinfo{volume}{8}},
  \bibinfo{pages}{2489--2494} (\bibinfo{year}{2012}).

\bibitem{schumacherEngineeredAntifoulingMicrotopographies2007}
\bibinfo{author}{Schumacher, J.~F.} \emph{et~al.}
\newblock \bibinfo{title}{Engineered antifouling microtopographies
  \textendash{} effect of feature size, geometry, and roughness on settlement
  of zoospores of the green alga {{{\emph{Ulva}}}}}.
\newblock \emph{\bibinfo{journal}{Biofouling}} \textbf{\bibinfo{volume}{23}},
  \bibinfo{pages}{55--62} (\bibinfo{year}{2007}).

\end{thebibliography}

\newpage

\begin{figure*}[htb]
\begin{center}
\includegraphics[width=0.7\textwidth]{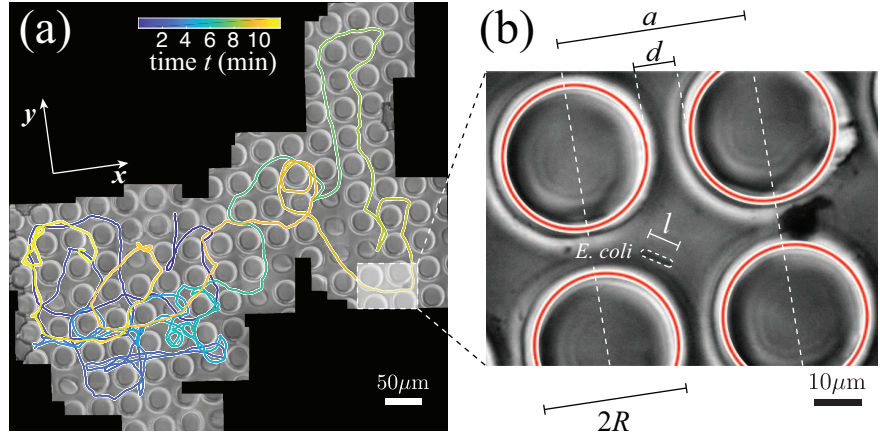}
\caption{Simultaneous large-scale and high-resolution study of bacterial transport in a micropillar array through active tracking and image stitching. (a) The trajectory of a single {\ec} (tracked up to ${\sim}1$ mm in distance and ${\sim}10$ minutes in time) was reconstructed to characterize its long-term transport. The trajectory is color coded in time. (b) The high resolution that was preserved in each original frame in (a) provided the detailed cell and pillar geometry. The pillars were $R=15$ $\mu$m in radius, arranged in a square lattice with a lattice constant of $a=40$ $\mu$m, and a gap of $d=a-2R=10$ $\mu$m between adjacent pillars. The tracked bacterium (in a dashed contour) was $l\approx 6$ $\mu$m in length. 
\label{fig:intro}}
\end{center}
\end{figure*}

\begin{figure}[htb]
\begin{center}
\includegraphics[width=0.7\textwidth]{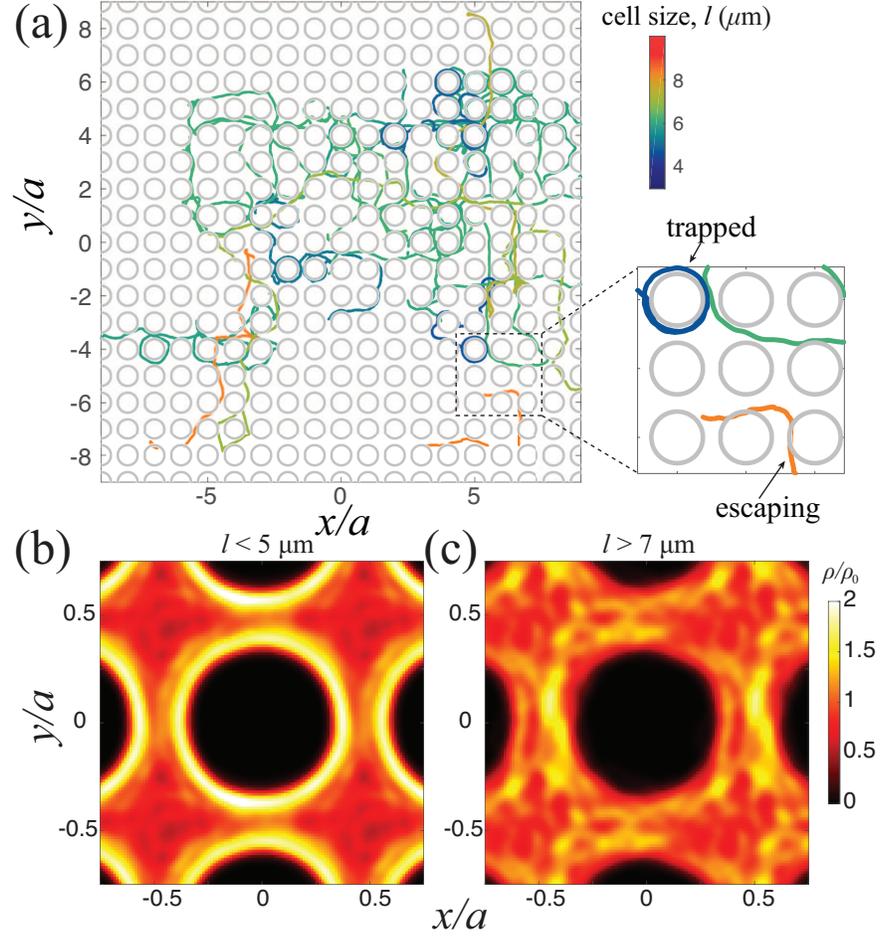}
\caption{
Cell size-dependent trapping and escaping effects. (a) Multiple trajectories for different individuals (color coded by cell length $l$) are mapped to the same pillar array. The shorter cells tend to circulate around the pillars while the longer cells tend to navigate between the pillars, indicating the distinct trapped and escaping mechanisms, respectively (highlighted in the inset). (b) Probability distribution $\rho$ of the shorter cells ($l<5$ $\mu$m), normalized by a uniform density $\rho_0$, shows an effective enhancement at the pillar surface. (c) A similar plot for longer cells ($l>7$ $\mu$m) shows the opposite effect.
\label{fig:diffuse}}
\end{center}
\end{figure}

\begin{figure*}[htb]
\begin{center}
\includegraphics[width=0.7\textwidth]{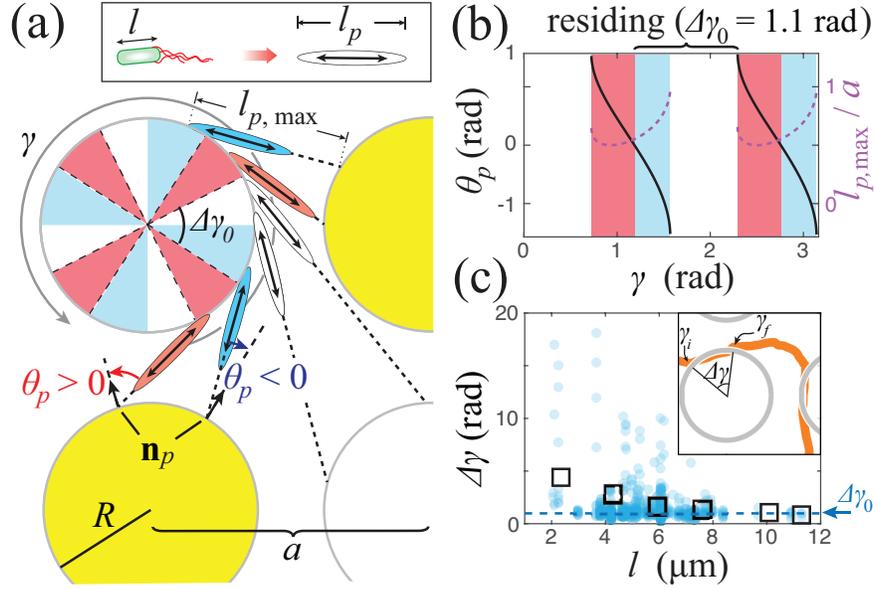}
\caption{
Lattice-constrained bacterial residency on the pillar surface. (a) A ``pusher'' representation of the bacterium (inset, with a cell body length $l$ and a total pusher length $l_p$) illustrates the geometric constraints for a bacterium circulating around a pillar (at angular position $\gamma$). An elongation of $l_p$ to $l_{p,\mathrm{max}}$ (the greatest possible $l_p$ without intersecting neighboring pillars) shows that the adjacent pillar (yellow) can either provide a positive (blue) or a negative (red) contribution to the circulation, demarcated by the orientation ($\theta_p$) of the pusher relative to surface normal of a neighboring pillar $\mathbf{n}_p$. This leads to periodically attractive ($\theta_p<0$ or without adjacent neighbors; blue or white) and repulsive ($\theta_p>0$; red) zones on a pillar. (b) A computation of $\theta_p$ (solid lines) and maximum $l_{p, \mathrm{max}}$ (dashed lines) give rise to a residency arc angle $\Delta\gamma_0=1.1$ rad (for $a=40$ $\mu$m, $R=15$ $\mu$m). (c) The residency arc angles $\Delta \gamma$ (squares), averaged over individual residency events (filled circles), decrease with increasing $l$ and eventually to values below the size of the attractive zone $\gamma_0$ (when $l\gtrsim 10$ $\mu$m), confirming suppressed circulation for longer cells. 
\label{fig:geomodel}}
\end{center}
\end{figure*}

\begin{figure}[htb]
\begin{center}
\includegraphics[width=0.7\textwidth]{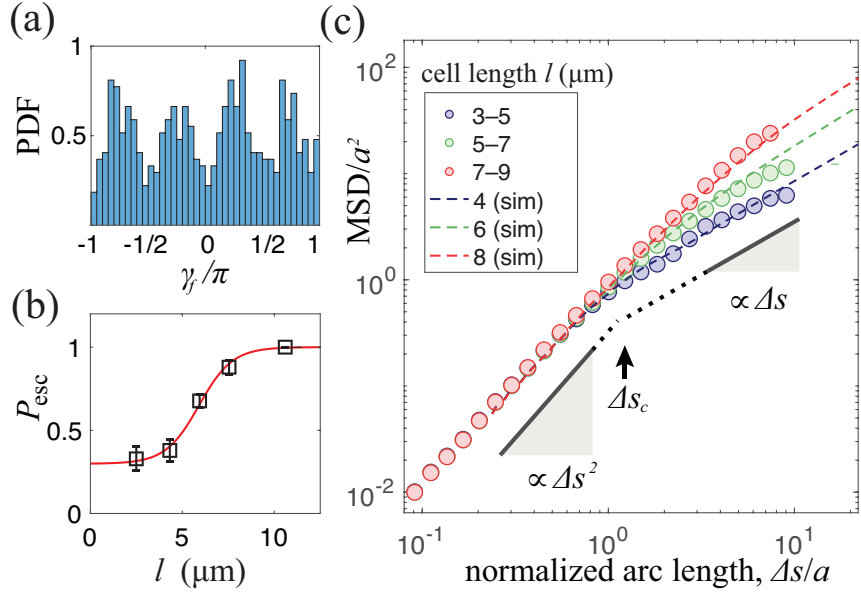}
\caption{Size-dependent escaping and global diffusivity. (a) The non-uniform distribution of angles of escaping ($\gamma_\mathrm{f}$, illustrated in Fig.~3c, inset) shows more probable escaping of bacteria along the diagonals (i.e., $k\pi/4$ with $k=\pm1$, $\pm 3$) of the square lattice, consistent with the locations of the repulsive zones.  
%Computed resident angles of a bacterium $\Delta \gamma$, assuming a Poisson process and a uniform escaping probability only at the repulsive zone $P_\mathrm{esc}$ (see SI), decrease monotonically with $P_\mathrm{esc}$. Shaded areas correspond to the standard deviation of the computed $P_\mathrm{esc}$ (with upper and lower quantiles treated separately).
(b) The escaping probabilities $P_\mathrm{esc}$ are calculated from bacterial trajectories for different cell lengths $l$ (open squares with error bars representing the standard errors).  
%The size-dependent escaping probability in the repulsive zone, $P_\mathrm{esc} (l)$ (open squares), is obtained by matching the computed $\Delta \gamma (P_\mathrm{esc})$ and experimentally measured $\Delta \gamma (l)$ (Fig.~3c). 
The solid curve corresponds to a fit with a hyperbolic tangent function. (c) The mean squared-displacement (MSD) as a function of the path length ($\Delta s$) exhibits a transition from a ballistic regime ($\mathrm{MSD}\propto \Delta s^2$) to a diffusive one ($\mathrm{MSD}\propto \Delta s$) for both experiments (circles) and numerical simulations (see SI) with the corresponding $P_\mathrm{esc}$ (dashed lines). The size of the ballistic regimes, depicted by a ballistic length $\Delta s_c$ (arrow), increase with $l$ or $P_\mathrm{esc}$, consistent with the geometry-induced escaping for longer cells. All lengths in trajectories are normalized by lattice size $a$.
 %The mean squared-displacement (MSD) as a function of the path length ($\Delta s$) for the trajectory a geometry-based model swimmer (with a escaping probability $P_\mathrm{esc}$) exhibits a transition from a ballistic regime ($\mathrm{MSD}\propto \Delta s^2$) to a diffusive one ($\mathrm{MSD}\propto \Delta s$) at a crossover length $\Delta s_c$ (inset). This $\Delta s_c$ increases with increasing $P_\mathrm{esc}$. (b) Experimental measurements (with dashed lines and shaded areas corresponding to the mean and standard deviations) show a similar dependency on cell lengths $l$. All lengths in trajectories are normalized by lattice size $a$. (c) A comparison of the experimentally measured residency arc angles $\Delta \gamma$ (circles) with the model predictions (dashed line) for various $l$ and $P_\mathrm{esc}$ provides a $P_\mathrm{esc}$ vs. $l$ function (inset). Here, $l$ is normalized by the averaged maximum pusher lengths $\bar{l}_p\equiv \left<l_{p, \mathrm{max}}\right>_\gamma$  constrained by geometry (Fig.~\ref{fig:geomodel}). (d) This $P_\mathrm{esc}(l)$ provides a prediction of the cell size ($l$) dependent transport characteristic $\Delta s_c$ (dashed line), which agrees well with the experiments (circles). The error bars and the shaded areas denote the standard deviations of the distributions of the relevant quantities in experiments and simulations, respectively. 
\label{fig:vsize}}
\end{center}
\end{figure}

\end{document}